\documentclass[conference]{IEEEtran}      

\usepackage[bookmarks=false]{hyperref} 
\hypersetup{draft}
\usepackage[utf8]{inputenc}
\usepackage{amsmath,amsfonts,amsbsy,amssymb}
\usepackage{mathabx}
\usepackage{mathrsfs}
\usepackage{tabularx}
\usepackage{graphicx}
\usepackage{cite}
\usepackage{wasysym}
\usepackage{multirow}
\usepackage{float}
\usepackage{color}
\usepackage{epstopdf}
\usepackage{mathtools}
\usepackage{graphics} 
\usepackage{graphicx}
\usepackage{siunitx}
\usepackage[nolist]{acronym}
\usepackage{siunitx}
\usepackage{eurosym}
\usepackage{footnote}

\definecolor{green1}{RGB}{70, 180, 80}

\title{Brain-to-brain Wireless Communication and Technologies Beyond \ac{5G}}

\author{Dick Carrillo$^{1}$, Renan Moioli$^{2}$, Pedro Nardelli$^{1}$\\
$^{1}$ School of Energy Systems, LUT University, Finland\\
$^{2}$ Digital Metropolis Institute, Federal University of Rio Grande do Norte, Brazil\\
}

\begin{document}
\maketitle
\thispagestyle{empty}
\pagestyle{empty}

\begin{abstract}
During the last few years, intensive research efforts are being done in the field of brain interfaces to extract neuro-information from the signals representing neuronal activities in the human brain. 
A recent development of these interfaces is capable of direct communication between animals' brains, enabling direct  brain-to-brain communication.
Although these results are new and the experimental scenario simple, the fast development in neuroscience, and information and communication technologies indicate the potential of new scenarios for wireless communications between brains.
Depending of the specific kind of neuro-activity to be communicated, the brain-to-brain link shall follow strict requirements of high data rates, low-latency, and reliable communication.
In this paper we highlight key beyond 5G technologies that potentially will support this promising approach.

\end{abstract}
\begin{acronym}
  \acro{1G}{first generation of mobile network}
  \acro{1PPS}{1 pulse per second}
  \acro{2G}{second generation of mobile network}
  \acro{3G}{third generation of mobile network}
  \acro{4G}{fourth generation of mobile network}
  \acro{5G}{fifth generation of mobile network}
  \acro{ARQ}{Automatic repeat request}
  \acro{ASIP}{Application Specific Integrated Processors}
  \acro{AWGN}{additive white Gaussian noise}
   \acro{BER}{bit error rate}
  \acro{BCH}{Bose-Chaudhuri-Hocquenghem}
  \acro{BRIC}{Brazil-Russia-India-China}
  \acro{BS}{base station}
  \acro{CDF}{Cumulative Density Function}
  \acro{CoMP} {cooperative multi-point}
  \acro{CP}{cyclic prefix}
  \acro{CR}{cognitive radio}
  \acro{CS}{cyclic suffix}
  \acro{CSI}{channel state information}
  \acro{CSMA}{carrier sense multiple access}
  \acro{DFT}{discrete Fourier transform}
  \acro{DFT-s-OFDM}{DFT spread OFDM}
  \acro{DSA}{dynamic spectrum access}
  \acro{DVB}{digital video broadcast}
  \acro{DZT}{discrete Zak transform}
  \acro{eMBB} {Enhanced Mobile Broadband}
  \acro{EPC}{evolved packet core}
  \acro{FBMC}{filterbank multicarrier}
  \acro{FDE}{frequency-domain equalization}
  \acro{FDMA}{frequency division multiple access}
  \acro{FD-OQAM-GFDM}{frequency-domain OQAM-GFDM}
  \acro{FEC}{forward error control}
  \acro{F-OFDM}{Filtered Orthogonal Frequency Division Multiplexing}
  \acro{FPGA}{Field Programmable Gate Array}
  \acro{FTN}{Faster than Nyquist}
  \acro{FT}{Fourier transform}
  \acro{FSC}{frequency-selective channel}
  \acro{GFDM}{Generalized Frequency Division Multiplexing}
  \acro{GPS}{global positioning system}
  \acro{GS-GFDM}{guard-symbol GFDM}
  \acro{IARA}{Internet Access for Remote Areas}
  \acro{ICI}{intercarrier interference}
  \acro{IDFT}{Inverse Discrete Fourier Transform}
  \acro{IFI}{inter-frame interference}
  \acro{IMS}{IP multimedia subsystem}
  \acro{IoT}{Internet of Things}
  \acro{IP}{Internet Protocol}
  \acro{ISI}{intersymbol interference}
  \acro{IUI}{inter-user interference}
  \acro{LDPC}{low-density parity check}
  \acro{LLR}{log-likelihood ratio}
  \acro{LMMSE}{linear minimum mean square error}
  \acro{LTE}{Long-Term Evolution}
  \acro{LTE-A}{Long-Term Evolution - Advanced}
  \acro{M2M}{Machine-to-Machine}
  \acro{MA}{multiple access}
  \acro{MAR}{mobile autonomous reporting}
  \acro{MF}{Matched filter}
  \acro{MIMO}{multiple-input multiple-output}
  \acro{MMSE}{minimum mean square error}
  \acro{MRC}{maximum ratio combiner}
  \acro{MSE}{mean-squared error}
  \acro{MTC}{Machine-Type Communication}
  \acro{NEF}{noise enhancement factor}
  \acro{NFV}{network functions virtualization}
  \acro{OFDM}{Orthogonal Frequency Division Multiplexing}
  \acro{OOB}{out-of-band}
  \acro{OOBE}{out-of-band emission}
  \acro{OQAM}{Offset Quadrature Amplitude Modulation}
  \acro{PAPR}{Peak to average power ratio}
  \acro{PDF}{probability density function}
  \acro{PHY}{physical layer}
  \acro{QAM}{quadrature amplitude modulation}
  \acro{PSD}{power spectrum density}
  \acro{QoE}{quality of experience}
  \acro{QoS}{quality of service}
  \acro{RC}{raised cosine}
  \acro{RRC}{root raised cosine}
  \acro{RTT} {round trip time}  
  \acro{SC}{single carrier}
  \acro{SC-FDE}{Single Carrier Frequency Domain Equalization}
  \acro{SC-FDMA}{Single Carrier Frequency Domain Multiple Access}
  \acro{SDN}{software-defined network}
  \acro{SDR}{software-defined radio}
  \acro{SDW}{software-defined waveform}
  \acro{SEP}{symbol error probability}
  \acro{SER}{symbol error rate}
  \acro{SIC}{successive interference cancellation}
  \acro{SINR}{signal-to-interference-and-noise ratio }
  \acro{SMS}{Short Message Service}
  \acro{SNR}{signal-to-noise ratio}
  \acro{STC}{space time code}
  \acro{STFT}{short-time Fourier transform}
  \acro{TD-OQAM-GFDM}{time-domain OQAM-GFDM}
  \acro{TTI}{time transmission interval}
  \acro{TR-STC}{Time-Reverse Space Time Coding}
  \acro{TR-STC-GFDMA}{TR-STC Generalized Frequency Division Multiple Access}
  \acro{TVC}{ime-variant channel}
  \acro{UFMC}{universal filtered multi-carrier}
  \acro{UF-OFDM}{Universal Filtered Orthogonal Frequency Multiplexing}
  \acro{UHF}{ultra high frequency}
  \acro{URLL}{Ultra Reliable Low Latency}
  \acro{V2V}{vehicle-to-vehicle}
  \acro{V-OFDM}{Vector OFDM}
  \acro{ZF}{zero-forcing}
  \acro{ZMCSC}{zero-mean circular symmetric complex Gaussian}
  \acro{W-GFDM}{windowed GFDM}
  \acro{WHT}{Walsh-Hadamard Transform}
  \acro{WLAN}{wireless Local Area Network}
  \acro{WLE}{widely linear equalizer}
  \acro{WLP}{wide linear processing}
  \acro{WRAN}{Wireless Regional Area Network}
  \acro{WSN}{wireless sensor networks}
  \acro{ROI}{return on investment}
  \acro{NR}{new radio}
  \acro{SAE}{system architecture evolution}
  \acro{E-UTRAN}{evolved UTRAN}
  \acro{3GPP}{3rd Generation Partnership Project }
  \acro{MME}{mobility management entity}
  \acro{S-GW}{serving gateway}
  \acro{P-GW}{packet-data network gateway}
  \acro{eNodeB}{evolved NodeB}
  \acro{UE}{user equipment}
  \acro{DL}{downlink}
  \acro{UL}{uplink}
  \acro{LSM}{link-to-system mapping}
  \acro{PDSCH}{physical downlink shared channel}
  \acro{TB}{transport block}
  \acro{MCS}{modulation code scheme}
  \acro{ECR}{effective code rate}
  \acro{BLER}{block error rate}
  \acro{CCI}{co-channel interference}
  \acro{OFDMA}{orthogonal frequency-division multiple access}
  \acro{LOS}{line-of-sight}
  \acro{VHF}{very high frequency}
  \acro{pdf}{probability density function}
  \acro{ns-3}{Network simulator 3}
  \acro{Mbps}{mega bits per second}
  \acro{IPC}{industrial personal computer}
  \acro{RSSI}{received signal strength indicator}
  \acro{OPEX}{operational expenditures}
  \acro{H2H}{human-to-human}
  \acro{DOD}{depth of discharge}
  \acro{ADC}{analog-to-digital converter}
  \acro{FFT}{Fourier fast transform}
  \acro{DAC}{digital-to-analog converter}
  \acro{IFFT}{inverse Fourier fast transform}
  \acro{DC/DC}{Direct-to-Direct current converter}
  \acro{BBI}{brain-to-brain interface}
  \acro{BBC}{brain-to-brain communication}
  \acro{SON}{self-organized networks}
  \acro{NOMA}{non-orthogonal multiple access}
\end{acronym}

\renewcommand\IEEEkeywordsname{Keywords}
\section{Introduction}
In \ac{BBC} the information is obtained by interfaces that combine neurostimulation, neuroimaging, and neuromodulation methods to obtain and deliver information between brains, allowing direct brain-to-brain communication. 
In very basic terms, information is extracted from the neural signals of a sender brain, digitized, and then delivered  to a receiver brain. Importantly, electrical or behavioral feedback improve performance of both subjects.
Early  interest  in \ac{BBC} came  from  the  potential for expanding  human  communication  and  social interaction capabilities \cite{ref_brain_1, ref_brain_2,pais2013brain,pais2015building}.
BBC is still in its infancy and lack several key features of real-world human communication. For example, subject interactivity has been minimal.

On a parallel front, new advances in \ac{BBC} have been designed to support communication for more than two human subjects. 
One possible scenario is composed by a communication structure that supports two Senders and one Receiver.
However, based on authors statement, it can be readily scaled up to include multiple Senders.
%
%
Additionally, researchers are currently exploring the use of novel techniques as functional magnetic resonance imaging (fMRI) that will increase the bandwidth of data transmitted over \ac{BBC} interfaces \cite{ref_b2b_com_1,ref_b2b_com_2}. 
%

Nevertheless, the recent \ac{BBC} results are still characterized by restricted information with low data rates.
In this sense, we would like here to point out potential requirements for novel wireless communication scenarios \cite{ref_wireless_b2b} where different nodes inside a specific wireless coverage are interchanging information through a variety of communication mechanisms (either half-duplex or full-duplex communication).
Besides, a variety of communication network topologies shall be driven by specific applications (from simple commands to a more complex conversation).
In this context, new challenges for the wireless network need to be addressed.
Specially, at the initial stages, an important requirement will be related to low latency with high reliability for low data rate.
However, as the technology matures, the network should support higher data rates as a future evolution of \ac{BBC} applications.
%

It is not difficult to realize the potential applications that \ac{BBC} will introduce in the society.
For example, the rehabilitation of persons suffering from poor motor conditions is one of the core drivers that motivate research in this field.
It is easy to see that not only health applications would benefit from the development of \ac{BBC}  but also 
industrial application in robotics and automation.
Entertainment and social relations may also become scenarios where \ac{BBC} could be the next (revolutionary) step after human-computer interfaces based on virtual/augmented reality and social media interactions. 

The role of technologies beyond \ac{5G} is important in this novel application for wireless networks.
In this short paper, we introduce a vision of future wireless networks that will potentially support \ac{BBC}.
This includes the network architecture, some potential applications, and its possible impairments, which is the focus of Section II.
In Section III, we listed some requirements and technologies that will enable \ac{BBC}.
Finally, a brief summary is done in Section IV.

\section{Network Architecture for Brain-to-brain Communication for Beyond 5G Technologies}
\begin{figure}[ht]
   \centering
   \includegraphics[width=\columnwidth]{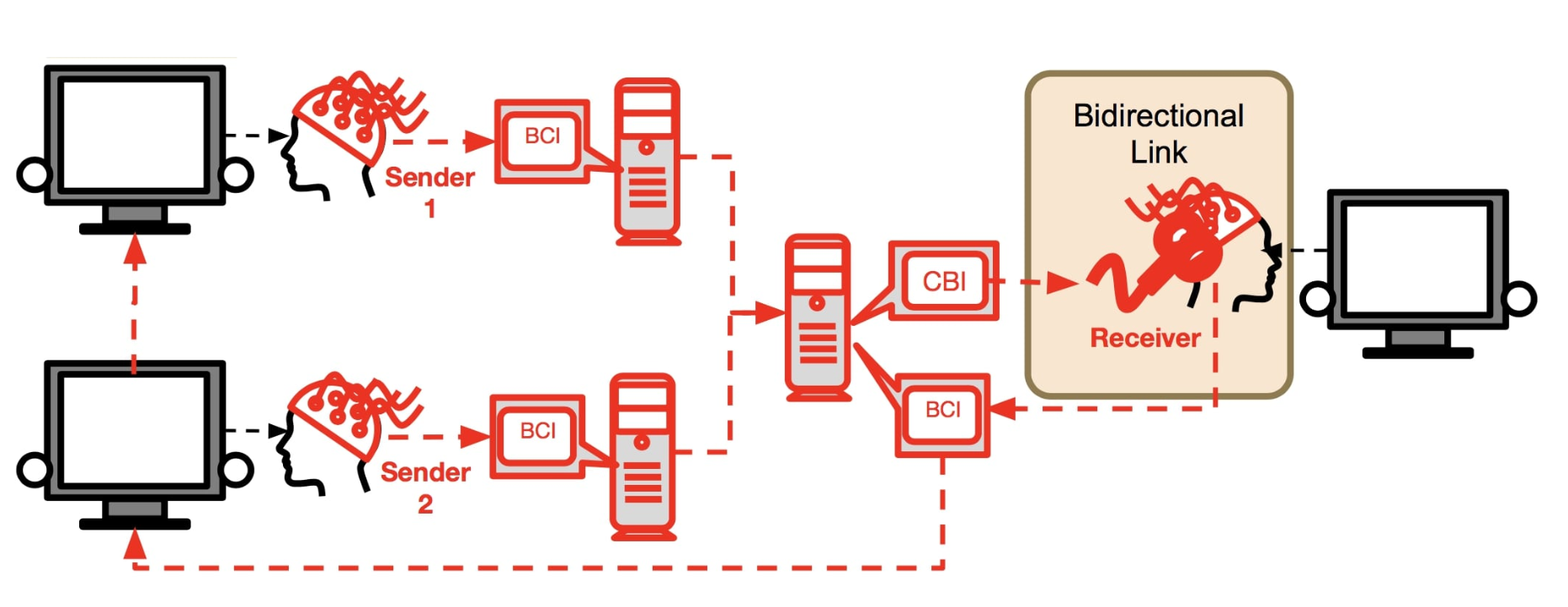}
   \caption{Architecture of the brain-to-brain communication developed in \cite{ref_b2b_com_1}. 
   }
   \label{fig:b2b_wired}
\end{figure}

A basic network architecture developed by early research results in \ac{BBC} is shown in Fig. \ref{fig:b2b_wired}.
Here is possible to check that actual prototypes use personal computers with high processing capacity and the communication channel is a wired link to optimize the network performance.
In wireless systems beyond \ac{5G}, one important requirement should be related to optimize latency/jitter and reliability using \ac{URLL} methodologies. 
Other important requirement is that wireless network should guarantee a true quality of service, even when the network considers simultaneously communication with high interference scenarios.

Based on the fact that \ac{BBC} is composed by a variety of topologies (point-to-point, point-to-multipoint) using a diversity of data propagation over the network (unicast, multicast, broadcast), one important requirement is flexibility in the radio resource management to optimize wireless channel resources.
In our vision, the first stage application of beyond \ac{5G} technologies in \ac{BBC} should consider a limited network coverage with capacity to apply radio resource management policies (as cellular network does) but without the dependency of a backhaul.
%
%
A mix of cellular network features with the practicality of non-licensed networks as Wi-Fi will converge to a network architecture that will depend strongly of \ac{SON} features to optimize the wireless system performance.
It is also important to mention that the concept of micro-operator with local licenses would be a potential solution.
\section{Beyond 5G Requirements and Enablers of  Brain-to-brain communication}
\subsubsection{\ac{GFDM}}
It is a flexible waveform that is a generalization of the popular \ac{OFDM} \cite{ref_GFDM_dick}. 
This flexibility enables the possibility to support different data rates and different latency values that is possible because \ac{GFDM} is a generalization of \ac{OFDM} with many optional configurations of subcarriers and subsymbols.
Other important advantage is that in the latest years many research investment was done in this waveform to improve \ac{GFDM} encoder performance.

\subsubsection{Non-Orthogonal Multiple Access}
Scenarios with high user density decrease the performance of the system.
However, when \ac{NOMA} is used, many research results have shown that the capacity network could be improved \cite{ref_noma_3}.
In the case of beyond \ac{5G}, we expect to apply hybrid solutions using \ac{NOMA} methodologies to support the various scenarios of \ac{BBC}.

\subsubsection{Artificial Intelligence}
Digital signal processing is a key tool in most of wireless communication technologies.
However, artificial intelligence techniques are gaining popularity, even in detection, channel estimation, radio resource scheduling, and power optimization \cite{ref_intel_artif}.
So, it is expected that popularity of artificial intelligence will provide a positive impact of \ac{BBC}, specially to optimize dynamically the network performance.
%
%
\subsubsection{\ac{URLL} in mmWaves}
The cellular generation \ac{5G} has two important key technologies to support independently industry requirements, they are \ac{URLL} and mmWaves (millimeter Waves).
However, to support \ac{BBC} it will be necessary that both technologies will join features in order to enable high level quality of service requirements.
\subsubsection{Local Core Network}
With the improvement of processors and hardware, it will be possible in beyond \ac{5G} technologies to run local core networks without the necessity to operate with a backhaul connectivity, as is used today.
This feature, will have strong consequences in the business model and application of these future networks \cite{5G_model_iran}.
This independence does not imply that features related to network management will be neglected. 
However, this networks will gain the popularity of WiFi with the advantage to support advanced features.
\vspace{-.2cm}
\section{Conclusions}
We presented the path for beyond 5G technologies to support \ac{BBC} based on the similarity of requirements with other future wireless applications.
Most of these requirements are related to \ac{URLL} communication and simultaneously the network considers scenarios with a high user density.
The diversity of topologies and use cases imply that beyond \ac{5G} technologies should be provided with intelligence to optimize the communication network.
It implies that the diversity of scenarios should be optimized from the waveform flexibility to the application of artificial intelligence in the radio resource management.
%
\vspace{-.2cm}
\bibliographystyle{IEEEtran}
\bibliography{main.bib}
\end{document}